\documentclass[a4paper,12pt]{article}
\usepackage[T1]{fontenc} 
\usepackage{graphicx}
\usepackage{hyperref}
\hypersetup{
    colorlinks=true,
    allcolors=blue}
\usepackage{pifont}
\usepackage{cancel}
\usepackage{pifont}
\usepackage{changes}
\usepackage{mathalfa,amssymb}
\usepackage{bbm}
\usepackage{bbold}
\usepackage{amsmath}
\usepackage{amsthm}
\usepackage{xcolor}
\usepackage[english]{babel}
\usepackage{multirow}
\usepackage[toc,page]{appendix}
\usepackage{hyperref}
\usepackage[margin=0.85in]{geometry}

\begin{document}
\begin{titlepage}
   \begin{center}
       \vspace*{3.5 cm}
        \LARGE
       \textbf{Asymptotic symmetries at null-infinity for the Rarita-Schwinger field with magnetic term\\}
       \vspace{1.5cm} 
       \large
       Bilyana L. Tomova \\
       \large
       \vspace{1.5cm}
        \textit{DAMTP, Centre for Mathematical Sciences, University of Cambridge,\\ Wilberforce Road, Cambridge,  CB3 0WA, U.K}
\vspace{1.5cm}

\end{center}

\abstract{In this paper we study the magnetic charges of the free massless Rarita-Schwinger field in four dimensional asymptotically flat space-time. This is the first step towards extending the study of the dual BMS charges to supergravity. The magnetic charges appear due to the addition of a boundary term in the action. This term is similar to the theta term in Yang-Mills theory. At null-infinity an infinite dimensional algebra is discovered, both for the electric and magnetic charge.}
 
\end{titlepage}
\flushbottom

\newpage

\tableofcontents

\newpage

\section{Introduction}
\label{sec:intro}

Asymptotic symmetries play a central role in the understanding of gauge theories. Gauge transformations can be divided into two important categories. On the one hand, we have the small gauge transformations, that are just redundancies in the description of the theory. On the other, there are the large gauge transformations, that transform the field to a physically inequivalent state. In this sense they are true symmetries of the theory and one can define their corresponding charges via Noether's theorem. These charges like the ADM mass in gravity, or the electric charge in the theory of electromagnetism, are defined as an integral over the boundary of some Cauchy slice. 

The first studied examples in asymptotically flat space-times were on space like slices, whose boundary is spatial infinity. However, it turned out that doing the analysis on a null surface leads to a more interesting and rich symplectic structure. In the case of four dimensional gravity, this work was pioneered by Bondi, Metzner and Sachs \cite{1}, \cite{7}. They found that the asymptotic symmetry group of asymptotically flat space-times is much larger than the naively expected Poincar\'{e} group. This symmetry group also gave rise to an infinite number of conservation laws, which can be roughly interpreted as the conservation of different modes of the mass and angular momentum flux. Later, it was proven that the BMS group is also a symmetry of the gravitational scattering matrix \cite{7} - the Ward identities of the corresponding conserved charges are equivalent to soft gravitational theorems \cite{8} (for a full review see \cite{20}). Therefore, asymptotic symmetries are very important for the quantization of the theory. Similar results have been found in other gauge theories \cite{9}, \cite{10}, \cite{11}, \cite{19}.

All this work has been insightful and fruitful. However, although the usual charges are very well understood, the dynamics of the elusive dual charges is still smewhat a mystery. In the case of electromagnetism, it was long believed that magnetic monopoles do not exist. That was until Dirac managed to construct a magnetic monopole from a carefully selected couple of gauge fields, that are each singular along a line starting from the origin. Using the ambiguity of the gauge connection, these singularities were avoided. More recently a quantization of the Maxwell theory was achieved with explicit electromagnetic duality and the corresponding duality charges were constructed \cite{23}. For the Einstein theory of gravity, the equivalent of this exotic solution is the Taub-NUT metric \cite{12}, \cite{13}. Indeed in \cite{14} it was shown that, the free data for the Einstein equations, written in Bondi gauge, contains fields that resemble Maxwell gauge fields. Choosing these gauge fields to be the Dirac monopole, one obtains the Taub-NUT metric. What is even more interesting is that, just as the usual NUT charge is the dual of the Bondi mass, an infinite number of charges, dual to the BMS charges, have been found \cite{15}. More recently their existence was proven rigorously \cite{16} in the Hamiltonian formalism of General Relativity. The correct way to achieve this is to add a topological term to the usual Einstein-Palatini action. While this term does not change the equations of motion, it does lead to a different symplectic structure of the theory and therefore to additional Hamiltonian charges. It is like the $\theta $ term in Yang-Mills theory.

The goal of this paper is to make the first step towards extending the study of dual charges to supergravity. Here we study the magnetic charges of the massless Rarita-Schwinger field in four-dimensional space-time on a fixed background. The asymptotic symmetries of the field in four dimensions have already been studied to some extend at null-infinity \cite{5} and spatial infinity \cite{2}. Here we do a similar analysis. We employ the covariant phase formalism (for a good review see \cite{17}, \cite{18}), to study the usual Rarita-Schwinger action plus a new boundary term, that will give rise to the new magnetic charges. Our study will differ from the previous ones in the choice of boundary conditions. We carefully selected them to allow for a finite and non-degenerate symplectic form and an infinite number of finite conserved Noether charges.

\section{Asymptotic behaviour}

Before introducing the action it is wise to first describe the setting. The Rarita-Schwinger field will be put on a fixed asymptotically flat four-dimensional space-time with vanishing Ricci tensor. Written in the Bondi gauge, this metric is \cite{1},

\begin{flalign}
 ds ^2 =  - e ^{2\beta}f^2 du ^2 - 2 e^{2\beta} dudr +g_{AB} (dx ^{A} - U^A du )(dx ^{A} - U^A du)
\end{flalign}
where the transverse part of the metric is written in the form $g_{AB} = r^2 h_{AB}$. The boundary conditions on the components of the metric are,

\begin{flalign}
& h _{A B}=  \gamma _{AB} +\frac{C _{AB}}{r}  + O(r^{-2})\\
&f ^2 (u, r, x^A)= 1 - \frac{2 M (u, x^A )}{r} +O(r^{-1}) \quad e ^{2\beta } =1+O (r^{-2} ) \quad U^A =- \frac{1}{2r^2} D_B C^{AB} + O(r^{-3})
\end{flalign}
where only the relevant components for the subsequent calculations are shown. The symbol $D_B$ is the covariant derivative with respect to the metric on the unit two-sphere $\gamma _{AB}$. The term $M(u, x^A)$ appearing is the Bondi mass. Unlike the ADM mass, it can depend on time and the angular coordinates. Furthermore, the subleading term $C_{AB}$  describes gravitational waves. There is an additional gauge freedom, that is used to set $\partial _r \det \left( \frac{g_{AB}}{r^2} \right) = 0 $. This implies that $\gamma ^{AB} C_{AB} = 0$, leaving out two degrees of freedom of this tensor, corresponding to the two polarizations of the gravitational wave.

In order to be able to talk about spinor vectors on this space-time, we need to introduce frame fields $e^a _{\mu }$,

\begin{flalign}
g ^{\mu \nu} e^a _{\mu } e^b _{\nu}  = \eta ^{ab}
\end{flalign}
where $\eta ^{ab}$ is the standard Minkowski metric with signature $(-, + , +, +)$.  An equivalent choice for a vierbein basis, would be a Lorentz transformation of the one, written above.The space-time is without torsion so the first structure equation is

\begin{flalign}
de^a + \omega ^a \,_{b} \wedge e^b = 0
\end{flalign}

Explicitly the frame fields for the asymptotically flat space-time can be chosen to be:

\begin{flalign}
&e^0 = \frac{e^{\beta }}{f}dr + e^{\beta } f du\quad 
e^1=  \frac{e^{\beta }}{f}dr  \\
&e ^i =  r E ^i _A \left ( dx^A   - U^A du \right) \quad E ^i _A  E^j _B \delta _{ij} =h _{AB} \quad i, j \in \{2, 3 \}
\end{flalign}

The expressions for the components of the spin connection $\omega $ are tedious and therefore, are put in the appendix.

Just as we did for the metric, we assume that the spinor field is analytic in \protect $1/r$, and its asymptotic behaviour at large  \protect $r$ is, 

\begin{flalign}
\psi _A = \psi _A ^{(0)} (x^A, u ) + O(r^{-1} ) \quad \psi _{u } = \frac{\psi _u ^{(-1)} (x^A, u )}{r} + O(r^{-2} ) \quad \psi _{r } = \frac{\psi _r ^{(-1)} (x^A, u)}{r} + O(r^{-2})
\end{flalign}

This describes the behaviour of the fields at $\mathcal{I}^+$ and it will be important when studying the Noether charges of the theory. It will ensure that these charges along with the energy and momentum are finite. The latter statement is tangential to the current discussion and is therefore shown in the appendix \ref{T}. The full set of diffeomorphism charges for the  $\mathcal{N} = 1$ supergravity theory is studied in an upcoming paper. Notice that the above conditions are different than the boundary conditions, proposed by \cite{5}, by order of $r$ in each component. Furthermore, they differ from \cite{2}, where the leading order of $\psi _A $ was pure gauge. 

We also need to fix the behaviour of the field at large $u$. We demand that,

\begin{flalign}
\lim _{u \to + \infty} \psi _{\mu } = \lim _{u \to - \infty} \psi _{\mu } < \infty
\end{flalign} 
where $< \infty $ indicates that the limit is finite. The above statement simply means that whatever $u$ dependence the field has, it vanishes at large  $|u|$.

We now move to think about what variations are allowed. We divide them into two categories - gauge and not gauge transformations. Our condition on the allowed variations is that the variational principle is well-defined (which is shown in the next section) and that the symplectic form is finite. For generic non-gauge variations, we demand that they do not change the leading behaviour of the field.

\begin{flalign}\label{10}
\delta \psi _A = O(r^{-1}) \quad \delta \psi _u = O(r^{-2}) \quad \delta \psi _A = O(r^{-2}) 
\end{flalign}

On the other hand, gauge transformations act on a specific way on the field, that allows for more flexibility. In particular the gauge transformation of the Rarita-Schwinger field is,

\begin{flalign}
\psi _{\mu} \rightarrow \psi _{\mu} + \nabla _{\mu } \epsilon
\end{flalign}
where \protect $ \epsilon $ is an anti-commuting spinor. One can think of this as the spinor analog of the \protect $U(1)$ gauge symmetry of electrodynamics. It turns out that the asymptotic behaviour for \protect  $\epsilon$ that gives finite, but non-vanishing charges is, 

\begin{flalign}
\epsilon  =  \frac{\epsilon ^{(-1)} (x^A, u)}{r} + O(r^{-2}) \quad \lim _{u \to \pm \infty } \epsilon < \infty 
\end{flalign}

We see that this is not compatible with \protect \eqref{10}, because \protect $\delta \psi _{u} = \nabla _{u } \epsilon = O(r^{-1})$. However, as we will see in the next sections, the variational principle and the symplectic form will still be well-defined, in the case, where the transformation is gauge.

These are all the assumptions we make about the asymptotic behaviour of the dynamical field \protect $\psi _{\mu }$ and the background field \protect $g_{\mu \nu }$. As already explained, the calculations done in this paper are well defined because of the above assumptions.

\section{The action }

The goal of this section is to introduce a new term to the action of the massless Majorana spinor vector field. The usual action is the Rarita-Schwinger, denoted in this paper by $S_{RS}$, while the new contribution is a boundary term, similar to the $\theta $ term in Yang-Mills theory and the Nieh-Yan and Pontryagin terms in the bosonic theory of gravity.

Without further ado, we introduce the action for the Rarita-Schwinger field with a magnetic term,
\begin{flalign}
\label{action}
S=S_{RS} + S_{\theta} = \int _{\mathcal{M}} d^4x  \,  e ^{\mu \nu \sigma \rho }  \bar{\psi } _{\mu } \gamma _5 \gamma _{\sigma }\nabla  _{\nu } \psi _{\rho } + \mathbb{\theta _{\ast} } \int _{\mathcal{M}} d^4 x \,  e ^{\mu \nu \sigma \rho } \bar{\psi } _{\mu }  \gamma _{\sigma } \nabla _{\nu } \psi _{\rho }
\end{flalign} 

The term $e^{\mu \nu \rho \sigma}$ is the alternating symbol and it obeys $e^{ur\theta \phi} =1$ and $\gamma _{\mu} = e^a _{\mu} \gamma _a$, where $a$ is a spinor index.  At first glance it doesn't seem like the added term is topological. However, we recall that a Majorana spinor obeys $\bar{\psi }  = \psi ^{\dagger } \gamma ^0=\psi ^T C$, where $C$ is the charge conjugation matrix. This matrix is antisymmetric and has the following conjugation property with the gamma matrices,  $\gamma _{\mu } ^T = -C \gamma _{\mu } C^{-1}$. Thanks to this, one can show that $\bar{\psi } _{\mu } \gamma _{\nu }\psi _{\rho} =  - \bar{\psi } _{\rho } \gamma _{\nu } \psi _{\mu}$ and more importantly that $ \bar{\psi } _{\mu } \gamma _{\nu } \nabla _{\sigma} \psi _{\rho} =  -  \left(\nabla _{\sigma }\bar{\psi } _{\rho } \right) \gamma _{\nu } \psi _{\mu}$. This allows to re-write the action as,

\begin{flalign}
S = \int d^4 x e ^{\mu \nu \sigma \rho } \,  \bar{\psi } _{\mu } \gamma _5 \gamma _{\sigma }\nabla _{\nu } \psi _{\rho } + \frac{\mathbf{\theta  _{\ast}}}{2} \int d^4 x \, \nabla _{\nu } (e ^{\mu \nu \sigma \rho } \bar{\psi } _{\mu }  \gamma _{\sigma }  \psi _{\rho } )
\end{flalign}

Now it is clear that the additional action term is a boundary term and therefore does not contribute to the equations of motion.

We now proceed to set the boundary conditions for the spinor field. The first thing to require is to have a well-posed variational principle. This means that on-shell $\delta S = 0$.

The variation of the action is given by,

\begin{flalign}
\delta S  =& \int d^4 x \, \left[ \frac{\partial L}{\partial \psi ^a } -\partial _{\mu } \left( \frac{\partial L}{\partial (\partial _{\mu } \psi ^a )} \right) \right]  \delta \psi ^a +\partial _{\mu } \left( \frac{\partial L}{\partial (\partial _{\mu } \psi ^a )} \delta \psi ^a \right)
\end{flalign}

The boundary term is the pre-symplectic potential, while the bulk one gives rise to the equations of motion.

\begin{flalign}
 &e ^{\lambda \nu \sigma \rho } \nabla _{\nu} \bar{\psi } _{\mu } \gamma _5 \gamma _{\sigma } =0
\end{flalign}

The equations of motion for the $\theta $ term is derived in the exact same way, only without the $\gamma _5$ matrix, namely $e ^{\lambda \nu \sigma \rho } \nabla _{\nu} \bar{\psi } _{\mu } \gamma _{\sigma } =0$. Since $\gamma ^5$ is invertible we have the following condition,

\begin{flalign}
e ^{\lambda \nu \sigma \rho } \nabla _{\nu} \bar{\psi } _{\mu } \gamma _5 \gamma _{\sigma } =0 \  \Leftrightarrow \ e ^{\lambda \nu \sigma \rho } \nabla _{\nu} \bar{\psi } _{\mu } \gamma _{\sigma } =0
\end{flalign}

Therefore, the additional term, has no contribution to the equation of motion.  We now turn to study the pre-symplectic potential in order to understand why it vanishes on the boundary.  Because the $\theta ^{\ast }$ term in the action can be re-written as a boundary term, its contribution to the pre-symplectic potential is straightforward,

\begin{flalign}
\delta S_b  = \frac{\mathbf{\theta  _{\ast}}}{2} \int  _{\partial \mathcal{M}}  d \Sigma _{\nu } \, e ^{\mu \nu \sigma \rho } \bar{\psi } _{\mu }  \gamma _{\sigma }  \delta \psi _{\rho }
\end{flalign}

The contribution from the usual Rarita-Schiwnger action is,

\begin{flalign}
\delta S_{RS} = \int _{\partial \mathcal{M}}  d \Sigma _{\nu }\, e ^{\mu \nu \sigma \rho } \bar{\psi } _{\mu }  \gamma _{\sigma }  \gamma _5 \delta \psi _{\rho }
\end{flalign}

Therefore on-shell the variation of the action is,

\begin{flalign}\label{theta}
\delta S _{\text{on-shell }}=  \int _{\partial  M } \, d\Sigma _{\nu } \,   \Theta ^{\nu }  =   \int _{\partial  M } \, d\Sigma _{\nu } \, e ^{\mu \nu \sigma \rho } \, \left( \bar{\psi } _{[\mu }  \gamma _{\sigma }  \gamma _5 \delta \psi _{\rho ] } + \frac{\mathbf{\theta  _{\ast}}}{2}  \bar{\psi } _{[\mu }  \gamma _{\sigma } \delta \psi _{\rho ] } \right)
\end{flalign}

The boundary of the manifold is the union of the limit of three different sequences of hypersurfaces \cite{24}. We evaluate the variation of the action at the surfaces defined by $u = \pm const $ and $r = const$ and then take the limit of the result as $u \to \pm \infty $ and $r \to \pm \infty $. Using the boundary conditions we have defined in the previous section is is not difficult to see that,

\begin{flalign}
\delta S _{\text{on-shell }} =  \lim _{r \to \infty } \int _{\Sigma _1} d \Sigma _r \Theta ^r +  \lim _{ u \to  + \infty } \int _{\Sigma _1} d \Sigma _u \Theta ^u +  \lim _{u \to  - \infty } \int _{\Sigma _1} d \Sigma _u \Theta ^u =0 
\end{flalign}

We see that on-shell the linear variation of the action, with respect to any variation of the field vanishes. For gauge transformations, there is even an easier way to see why they leave the action invariant on-shell.

\begin{equation} 
\begin{split}
S \rightarrow  \, S  +& \int d^4 x\, e ^{\mu \nu \sigma \rho }  \bar{\psi}_{\mu}\gamma _5 \gamma _{\sigma }\nabla  _{\nu } \nabla  _{\rho } \epsilon + \theta  _{\ast} \int d^4 x\, e ^{\mu \nu \sigma \rho } \bar{\psi} _{\mu}  \gamma _{\sigma } \nabla _{\nu} \nabla  _{\rho } \epsilon   \\
+&\int d^4 x \,e ^{\mu \nu \sigma \rho }  \nabla _{\mu} \bar{\epsilon}\gamma _5 \gamma _{\sigma } \nabla  _{\nu } \psi _{\rho} + \theta  _{\ast} \int d^4 x \, e ^{\mu \nu \sigma \rho } \nabla  _{\mu }\bar{\epsilon}  \gamma _{\sigma } \nabla _{\nu}  \psi _{\rho} = S
\end{split}
\end{equation}

The second line vanishes because of the equation of motion. In the first line we have we have, 

\begin{flalign}
&e ^{\mu \nu \sigma \rho } \gamma _{\sigma} [ \nabla _{\nu},  \nabla _{ \rho } ] \epsilon = \frac{1}{4}
  e ^{\mu \nu \sigma \rho } \gamma _{\sigma}R_{\nu \rho a b} \gamma ^{ab} \epsilon  =  \frac{\sqrt{g}}{4} R_{\nu \rho a b} \gamma ^{\mu \nu \rho }  \gamma ^{ab} \epsilon
\end{flalign}

This vanishes, because of the Bianchi identity of Riemann tensor and because the Ricci tensor vanishes. For the details of this calculations see \cite{21}. It is interesting that the invariance of the action under gauge transformations is independent of the boundary conditions for the field $\psi $. This shows yet again that gauge transformations are special.

\section{Symplectic structure}

In this section we study the symplectic structure of the theory. We start with a brief review of the covariant phase space formalism for gauge theories. 

The covariant phase space of a gauge theory is the space of solutions to the field equations with particular boundary conditions. The symmetries of the theory are gauge transformations, that preserve the gauge condition on the fields, and leave the action invariant. If the transformations are "large", they will have a non-vanishing conjugate charges. Without going into too much details, the recipe for calculating these charges is the following~\cite{17}. Firstly, the variation of the Lagrangian on shell is given by by a total derivative,

\begin{gather}\label{3.1}
\delta L (\phi, \delta \phi ) \approx d \Theta (\phi, \delta \phi )
\end{gather}

The boundary term $\Theta $ is called the pre-symplectic potential. The pre-symplectic current and pre-sympelctic form on phase space are defined respectively as,

\begin{flalign}\label{3.2}
& w (\phi, \delta _1 \phi, \delta _2 \phi ) = \delta _1 \Theta (\phi, \delta _2 \phi ) - \delta _2 \Theta (\phi, \delta _1 \phi )\\
& \tilde{ \Omega } (\phi, \delta _1 \phi, \delta _2 \phi ) = \int _{\Sigma } w (\phi, \delta _1 \phi, \delta _2 \phi )
\end{flalign}

We have used the $w$, to denote the pre-symplectic current, instead of the usual $\omega$, because $\omega $ is already reserved for the spin connection. In order to construct the symplectic form $\Omega$ from the pre-symplectic form, one should quotient out the degenerate directions $Y^b$ of the pre-symplectic form \cite{21}. These satisfy the property that $\Omega _{ab} Y^{a} X^b = 0$ for any $X^b$. Note that in this equation the latin letters are indices on the infinite dimensional phase space, not the spinor indices.

The Hamiltonian conjugate to a gauge transformation is defined as,

\begin{flalign} \label{dH}
& \cancel{\delta} H [\epsilon ] = \Omega _{ab} X^b [\epsilon ]= \int _{\Sigma } w (\phi, \delta _{ \epsilon} \phi, \delta \phi )
\end{flalign} 
where $X^b [\epsilon ]$ is a vector field in phase space, which connects field configurations, related by the gauge transformation generated by $ \epsilon$. The integral is taken over a Cauchy surface. Later we will pick it to be future null-infinity (plus future time-infinity). Equation \eqref{dH} can be rewritten as in \cite{3},

\begin{flalign}\label{H}
\cancel{\delta } H _{ \epsilon } = \int _{ \partial \Sigma } \delta \mathcal{Q} - I_{\epsilon} \cdot \Theta 
\end{flalign}
where $I_{\epsilon} = \epsilon \cdot \frac{\delta }{\delta \phi }$ and $\mathcal{Q}$ is the Noether charge. The corresponding Noether current is \cite{3}:

\begin{gather}
j=d\mathcal{Q} = \Theta - I _{ \epsilon} L
\end{gather}

In equation \eqref{dH}, I have deliberately used the symbol $\cancel{\delta } $ instead of $\delta $. This is because $\cancel{\delta } H _{\xi }$ need not be an exact one form on phase space. This can be due to two reasons - the transformation in question is not canonical, or because of the presence of flux of the charge $\mathbf{F}$ through null-infinity \cite{3}. In general, one makes the flux vanishing by fixing the right boundary conditions. Equipped with the tools of covariant phase space formalism, we now proceed to study the symplectic structure of our theory.

By using \eqref{3.2} and \eqref{theta} we can derive the pre-symplectic current,

\begin{flalign}
&w(\psi, \delta _1 \psi, \delta _2 \psi ) =\delta _1 \Theta (\psi, \delta _2 \psi) - \delta _2 \Theta (\psi, \delta _1 \psi)=\\ \label{3.12}
&\delta _1 \bar{\psi } _{[\mu } \gamma _{\sigma } \gamma _5 \delta _2 \psi _{\rho ] } - \delta _2 \bar{\psi } _{[\mu } \gamma _{\sigma } \gamma _5 \delta _1 \psi _{\rho ] } + \underbrace{\frac{\mathbf{\theta _{\ast}}}{2} \delta _1 \bar{\psi } _{[\mu } \gamma _{\sigma } \delta _2 \psi _{\rho ] } - \frac{\mathbf{\theta _{\ast}} }{2} \delta _2 \bar{\psi } _{[\mu } \gamma _{\sigma } \delta _1 \psi _{\rho ] } }_{=0}
\end{flalign}

Let's look at equation \eqref{3.12}. On one hand, from the properties of the Majorana spinors, discussed in the first section, we have $- \delta _2 \bar{\psi } _{[\mu } \gamma _{\sigma } \delta _1 \psi _{\rho ] } = + \delta _1 \bar{\psi } _{[\rho } \gamma _{\sigma } \delta _2 \psi _{\mu ] } $. On the other, the pre-symplectic current is antysiymmetrized over its indices. Therefore, the $\theta _{\ast}$ contribution to the pre-symplectic current vanishes.The symplectic structure of the theory is unaffected by the introduction of a boundary term in the action.

The pre-symplectic form on null-infinity is thus,

\begin{flalign}
\tilde{\Omega }= \int_{\mathcal{I} ^+} \, du d\theta d\phi \, \delta _1 \bar{\psi } _{[u } \gamma _{\theta} \gamma _5 \delta _2 \psi _{\phi] }
\end{flalign}

The corresponding Hamiltonian to a gauge transformation is,

\begin{flalign} 
&\cancel{\delta} H _{ \epsilon} = \int _{\Sigma } \delta \Theta (\psi, \nabla _{\mu } \epsilon ) - \delta _{ \epsilon} \Theta (\psi, \delta \psi)
\end{flalign}

\begin{equation}
\begin{split}
&\delta _{ \epsilon } \Theta (\psi, \delta \psi) _{\mu \sigma \rho } = \nabla _{[\mu } \bar{ \epsilon } \gamma _{\sigma } \gamma _5 \delta \psi _{\rho ] } +\frac{\theta _{\ast}}{2} \nabla _{[\mu } \bar{ \epsilon} \gamma _{\sigma } \delta \psi _{\rho ] } = \\
&\nabla _{[\mu } \left(\bar{ \epsilon } \gamma _{\sigma } \gamma _5 \delta \psi _{\rho ] } +\frac{\theta _{\ast}}{2} \bar{ \epsilon } \gamma _{\sigma } \delta \psi _{\rho ] } \right) + \bar{ \epsilon } \gamma _{\sigma } \gamma _5 \delta \nabla_{\mu} \psi _{\rho ] } + \frac{ \theta _{\ast}}{2} \bar { \epsilon } \gamma _{\sigma } \nabla _{\mu } \delta \psi _{\rho ] } =d\vartheta
\end{split}
\end{equation}

In the above expression we have written the variation of the symplectic potential along a gauge transformation as a total derivative, using the linearized equations of motion. This is a special property of gauge theories - the charges live on the boundary.

The above equations can be re-written more formally thanks to the identity,

\begin{flalign}
\iota _{\nabla _{\mu } \epsilon ^a } \frac{\delta }{\delta \psi ^a _{\mu }} (...) = d \left( \epsilon ^a \frac{\delta }{\delta \psi ^a _{\mu }} (...) \right) + \epsilon ^a \frac{\delta }{\delta \psi ^a _{\mu }} (d(...)) 
\end{flalign}

where "d" denotes the exterior derivative on the space-time manifold and $a$ is a spinor index. This identity is the similar of the Cartan's magic formula, but for fermions. Recalling that we defined $I_{\epsilon } = \epsilon ^a \frac{\delta }{\delta \phi ^a}$ this can be rewritten more compactly,

\begin{flalign}
\delta _\epsilon \Theta (\psi, \delta \psi) & = \iota _{\nabla _{\mu } \epsilon ^a } \Theta (\psi, \delta \psi) = d I_{\epsilon } \Theta (\psi, \delta \psi) + I_{\epsilon } d \Theta (\psi, \delta \psi)\\
d\vartheta & = d I_{\epsilon } \Theta (\psi, \delta \psi)
\end{flalign}

The first equality holds because of the way $\Theta $ is constructed from the field and its derivatives. More precisely it a n-2 form on space-time, constructed covariantly. Going back to $\cancel{\delta} H_e$ we have now,

\begin{flalign}
\begin{split}\label{noethercurrent}
\cancel{ \delta } H_{\epsilon } = &\int _{\Sigma } \delta \underbrace{( \Theta (\psi , \nabla _{\mu } \psi ) - \bar{\epsilon } \gamma _{\sigma } \gamma _5 \nabla _{\mu} \psi _{\rho ] } - \frac{\theta _{\ast}}{2} \bar{e} \gamma _{\sigma } \nabla _{\mu } \psi _{\rho ] } ) } _{ \text{Noether current } = \ast j} + d\vartheta \\
\approx & \int _{\Sigma } \delta (\nabla _{[ \mu }(\bar {\psi } _{\mu } \gamma _5 \gamma _{\sigma ] } \epsilon + \frac{\theta _{\ast}}{2} (\bar {\psi } _{\mu } \gamma _{\sigma ] } \epsilon )) + d\vartheta \\
= & \int _{\partial \Sigma } \bar{\epsilon } \gamma _{[\sigma } \gamma _5 \delta \psi _{\rho ] }
\end{split}
\end{flalign}

The $\theta$ contribution vanishes once again. At this point, one may think that the new action term is completely inconsequential. However, because of the peculiar properties of spinors, we will see that the Noether charges will be affected by the magnetic term, even though the Hamiltonian charges are not. This is the subject of the next section.

\section{Noether charge}

If a gauge theory, described by a Lagrangian, admits global symmetries we can apply the generalized Noether theorem \cite{17}. This theorem states that there exists a bijection between the gauge parameters, and the equivalence class of $d-2$ forms $\mathcal{Q}$ that are closed on shell. Two such forms are equivalent if on-shell they differ by a $d-3$ form of the type $dk$. The integral over the boundary of a Cauchy slice of these $d-2$ forms is the Noether charge. Looking back at  \eqref{H}, we see that the Noether charge is part of the integrable part of the Hamiltonian charge.

In this section we will calculate the Noether charge for the gravitino field with action \eqref{action}. The exterior derivative of the Noether charge density is given in \eqref{noethercurrent},

\begin{flalign}
\ast j=d \ast \mathcal{Q}=\Theta -I_{\epsilon } \ast L = \theta (\psi, \nabla _{\mu } \epsilon ) - \bar{\epsilon } \gamma _{[\sigma } \gamma _5 \nabla _{\mu} \psi _{\rho ] } - \frac{\theta _{\ast} }{2} \bar{\epsilon } \gamma _{[\sigma } \nabla _{\mu } \psi _{\rho ] }
\end{flalign}
where $\ast$ denotes the Hodge dual. The $I_{\epsilon } \ast L$ part of the equation vanishes, because of the equations of motion. This leads to,

\begin{flalign}
d \ast \mathcal{Q} [\epsilon ] _{\rho \mu \sigma } = &\bar{\psi } _{[\mu } \gamma _{\sigma } \gamma _5 \nabla _{\rho ] } \epsilon + \frac{\theta _{\ast}}{2} \bar{\psi } _{[\mu } \gamma _{\sigma }\nabla _{\rho ] } \epsilon \approx \nabla _{[ \rho } \bar{\psi } _{\mu } \gamma _{\sigma ] } \gamma _5 \epsilon + \frac{\theta _{\ast} }{2} \nabla _{[ \rho } \bar{\psi } _{\mu } \gamma _{\sigma ] } \epsilon \\
\ast \mathcal{Q} [\epsilon ] _{\mu \sigma } = & \bar{\psi } _{[\mu } \gamma _{\sigma ] } \gamma _5 \epsilon + \frac{\theta _{\ast} }{2} \bar{ \psi } _{[ \mu } \gamma _{\sigma ] } e\\
\mathbf{Q} [ \epsilon ] = & \int _{\partial \Sigma } \ast \mathcal{Q} [ \epsilon ] 
\end{flalign}

As promised, Noether charge, has a a non-vanishing magnetic contribution, despite the fact that the Hamiltonian charge does not. One can think of this peculiarity in the following way. The non-integrable part of $\cancel{\delta } H$ is $I_{\epsilon } \Theta $. It characterizes the flux of this charge through null-infinity. For the usual charge we have that $\delta \mathcal{Q} =- I_{\epsilon } \Theta $, which means that all of the charge is contained in the bulk. In contrast for the magnetic part, the relationship is $ \delta \mathcal{Q} _{\theta _{\ast}}= I_{\epsilon } \Theta _{\theta _{\ast}}$ - all of the magnetic charge leaks through infinity.

Now let's look at the algebra of the charges:,

\begin{flalign}
\begin{split}
[ \mathbf{Q} [ \epsilon _1], \mathbf{Q} [ \epsilon _2] ] \equiv & \frac{1}{2} (\delta _{ \epsilon _1 }\mathbf{Q} [ \epsilon _2] - \delta _{\epsilon _2}\mathbf{Q} [\epsilon _1] ) \\
= & \int _{\partial \Sigma } \bar{\psi } _{[\mu } \gamma _{\sigma ] } \gamma _5 [ \epsilon _1, \epsilon _2 ]+ \frac{\theta _{\ast} }{2} \bar{ \psi } _{[ \mu } \gamma _{\sigma ] } [ \epsilon _1 , \epsilon _2 ] \\ 
& + \int _{\partial \Sigma } \nabla _{[\mu } \bar{\epsilon }_1 \gamma _{\sigma ]} \gamma _5 \epsilon _2 +\frac{\theta _{\ast} }{2}\int _{\partial \Sigma } \nabla _{[\mu } \bar{\epsilon }_1 \gamma _{\sigma ]} \epsilon _2 - (\epsilon _1 \leftrightarrow \epsilon _2)\\
= & \underbrace{ \mathbf{Q} \left[ [\epsilon _1, \epsilon _2 ] \right] }_{=0}+ \text{central charge} 
\end{split}
\end{flalign}

The underlying algebra is abelian, so the result of the commutator is just the central charge. The central charge has two components- the usual one from the Rarita-Schwinger field and the magnetic one. In \cite{5}, it was shown that, if one defines a vector field as $\xi^{\mu} = \bar{ \epsilon } _1 \gamma ^{\mu } \epsilon _2$ and uses the linearized spin connection $\delta \omega$ in the calculation for the usual central charge, one obtains the super-translation charge, generated by $\xi ^{\mu}$. The $\theta $ contribution to the central charge vanishes because it is the integral of a total derivative on the sphere.

\section{Explicit form of the Noether charge}

In this section we compute the explicit form of the Noether charge. We start by fixing the gauge,

\begin{flalign}\label{gauge}
\gamma ^{\mu } \psi _{\mu } = 0
\end{flalign}

This leaves the the equations of motion in the form,

\begin{flalign}
\gamma ^{\mu }  \nabla_{\mu } \psi _{\nu } = 0
\end{flalign}

Explicitly the charge is,

\begin{flalign}
\begin{split}\label{Q}
\mathbf{Q} [\epsilon ] = &\int _{S^2 }  d\theta d\phi \, \bar{\psi } _{\phi}\gamma _5 \gamma _{\theta  } \epsilon  + \frac{\theta  _{\ast} }{2} \bar{\psi } _{\phi} \gamma _{\theta} \epsilon  - (\theta \leftrightarrow \phi )  \\
=&- \int _{S^2 }  d\theta d\phi \bar{\psi } _{A} \gamma ^A \gamma _5 \gamma _{\phi }\gamma _{\theta  } \epsilon  + \frac{\theta  _{\ast} }{2} \bar{\psi } _{A} \gamma ^A \gamma _{\phi} \gamma _{\theta} \epsilon 
\end{split}
\end{flalign}
where $S^2$ is the two sphere at null-infinity. In order to evaluate the charge we will need the expression for both the gauge spinor and the angular components of the Rarita-Schwinger field. Therefore we need to solve their equations of motion. We start by looking at the gauge spinor, which is simpler and its solution will help us with the subsequent calculations.

The gauge spinor satisfies the Dirac equation as imposed by the gauge condition \eqref{gauge},

\begin{flalign}
\begin{split}
& \gamma ^{\mu }  \nabla_{\mu } \epsilon   =0 \quad  \epsilon  = \sum _{n \geq 1 } \frac{ \epsilon ^{-n}}{r^n} \quad \gamma ^{\mu}  \nabla_{\mu } - \gamma ^r \partial _r= \sum _{n \geq 1 } \frac{v ^{-n}}{r^n}\\
& \gamma ^{\mu }  \nabla_{\mu } \epsilon  =  \sum_{n \geq 0 } \sum _{m=0} ^{m=n} \frac{\left(  \nabla^{-m} \epsilon  ^{m-n} +(m-n+1) \gamma ^{r-m} \epsilon ^{m-n+1}\right)}{r^n} =0
\end{split}
\end{flalign}

At first and second order in $r$ the equations of motion are the following,

\begin{flalign}
& n=1 :  \gamma ^{u 0} \partial _u \epsilon ^{-1}=0  \\\label{n1}
& n=2 :  \gamma ^{A(-1)} \partial _A \epsilon ^{-1}   + \frac{1}{2} \cot \theta \gamma ^2  \epsilon ^{-1} +\gamma ^{u 0}\partial _{u} M \epsilon ^{-1} +\gamma ^{u 0} \partial _u \epsilon ^{-2} =0 
\end{flalign}

The first equation leaves us with,

\begin{flalign}
&\epsilon ^{-1}=\rho(x^A) +  \gamma ^{u(0)} \beta (u, x^A) 
\end{flalign}

Notice that we can have a $u$ dependency, because  $\gamma ^{u(0)} \gamma ^{u(0)} = 0 $. Similarly to the gravitational case, working at null-infinity allows us to have more interesting dynamics. Using the representation of the gamma matrices, given in the appendix, we can write $\left( \gamma ^{u(0)} \beta  ^{-1} \right) ^T= \begin{pmatrix} \eta ,& \xi \end{pmatrix}$, where $\eta ^T=\begin{pmatrix} \eta ^1 , & \eta ^1 \end{pmatrix}$  and $\xi ^T= \begin{pmatrix} \xi _1  , & -\xi _1\end{pmatrix}$.We also have $ \left( \rho  ^{-1} \right)^T = \begin{pmatrix} \iota ,& \lambda \end{pmatrix} $. The equations for the chiral spinors decouple and will be solved separately .  We will proceed to treat the equations of $\rho$ and $\beta $  separately and in a different fashion. We assume separation of variables for the components of both spinors. Let's first look at $\rho$. It satisfies the time independent part of equation \eqref{n1}.

\begin{flalign}
&\gamma ^{A(-1)} \partial _A \rho   + \frac{1}{2} \cot \theta \gamma ^2  \rho = 0 
\end{flalign}

From here, we can extract the equations for the two components of the chiral spinor $\lambda$.

\begin{flalign}
& \partial _{\theta } (\lambda _1  + \lambda _2)- \frac{i}{\sin \theta } \partial _{\phi } (\lambda _1  + \lambda _2) + \frac{1}{2} \cot \theta  (\lambda _1  + \lambda _2)= 0\\
& \partial _{\theta } (\lambda _1  - \lambda _2)+ \frac{i}{\sin \theta } \partial _{\phi } (\lambda _1  - \lambda _2) + \frac{1}{2} \cot \theta  (\lambda _1  - \lambda _2)= 0
\end{flalign}

This is solved by $\lambda _1   - \lambda _2 =\frac{A}{\sqrt{\sin \theta }} e^{-im \phi }  \left( \cot \theta /2 \right) ^{m}$ and  $\lambda _1    +  \lambda _2 =  \frac{B}{\sqrt{\sin \theta }} e^{im \phi }  \left( \cot \theta /2 \right) ^{m}$. The same set of equations and solutions hold for $\iota$. This sets everything we need to know about $\rho$.

Now we move on to studying the equation for components of $\eta$ and $\xi$. 

\begin{flalign}
\begin{split}\label{5.20}
 & \partial _{\theta }\xi  _1 - \frac{i}{\sin \theta } \partial _{\phi } \xi   _1 + \frac{1}{2} \cot \theta \xi  _1 =-i \partial _u M(\lambda _1 - \lambda _2) - i \partial _u (\epsilon ^{-2} _3 - \epsilon ^{-2} _4)  \\
&- \bar{\eth } _{1/2} \xi _1  = -i \partial _u M(\lambda _1 - \lambda _2) - i \partial _u (\epsilon ^{-2} _3 - \epsilon ^{-2} _4)
\end{split}
\end{flalign}

\begin{flalign}
\begin{split}\label{5.21}
& \partial _{\theta }\eta _1 + \frac{i}{\sin \theta } \partial _{\phi } \eta  _1 + \frac{1}{2} \cot \theta \eta  _1 =   i \partial _u M(\lambda _1 - \lambda _2)  + 2 \partial _u (\epsilon ^{-2} _1 + \epsilon ^{-2} _2)  \\
&-\eth _{-1/2} \eta _1 =  i \partial _u M(\lambda _1 - \lambda _2)  + 2 \partial _u (\epsilon ^{-2} _1 + \epsilon ^{-2} _2)
\end{split}
\end{flalign}
 
The operators $\eth$ and $\bar{\eth }$ are defined in \cite{4}. The first thing we notice is that $\xi _1$ and $\eta _1$ should be expressed in terms of spherical harmonics of spin weight $1/2$ and $-1/2$ respectively. Furthermore, we observe that we have two equations and four unknown functions. Therefore we claim that $\xi _1$ and $\eta _1$ can be expressed as products of arbitrary functions of $u$ and a spin $\pm 1/2$ spherical harmonics. Then the above equations can be viewed as equations for $\epsilon ^{-2}$. Of course, $\epsilon ^{-2}$, will have another equation of motion, but in that $\epsilon ^{-3}$ will appear etc. We see that whatever choice we make for leading order component of the gauge spinor, all the subsequent components will adjust to accommodate it. Thus. we have a complete freedom for $\epsilon ^{-1}$. What is left now is to impose the Majorana condition $\left(\epsilon ^{-1} \right)^{ \dagger} \gamma ^0 = \left( \epsilon ^{-1} \right) ^T C$. 

The final piece of the puzzle we need is the behavior of the field itself. Its equations of motion at first and second order in $r$ are,

\begin{itemize}

\item{$\psi _A$}:
\begin{flalign}\label{uA}
O(1) : & \gamma ^{u(0)} \partial _u \psi _A =0 \\\label{A2}
O(r^{-1}) : &\gamma ^B \partial _B \psi _A  + \gamma ^1 \frac{1}{r}\psi _A + \frac{1}{2r} \cot \theta \gamma ^2 \psi _A -\frac{1}{r} \gamma ^r  \psi _A - \frac{r}{2} \gamma ^u g^{BE} \partial _u C_{EA} \psi _B\\
& -  \gamma ^C \Gamma ^B _{CA} \psi _B+ h _{AB} \gamma ^B \psi _{u} +\frac{1}{2} \partial _u C_{AB} \gamma ^B \psi _r +  \gamma ^{u(0)} \partial _u \psi _A  ^{-1} =0
\end{flalign}

\item{$\psi _u$:}
\begin{flalign}
O(r^{-1}) : & \gamma ^{u } \partial _{u } \psi _u  =0 \\ \label{u2}
O(r^{-2}): &  r  \gamma ^A \partial _A \psi _u  + \frac{1}{2} \cot \theta \gamma ^2 \psi _u +  \gamma ^u \partial _u M ( \psi _u - \psi _r )  +  \gamma ^{u } \partial _{u } \psi _u ^{-2} - r \gamma ^A   \partial _u C_{A} ^B \psi _B  =0
\end{flalign}

\item{$\psi _r$:}
\begin{flalign}
O(r^{-1}) :  &\gamma ^{u } \partial _{u } \psi _r   =0\\ \label{r2}
O(r^{-2}) : & r \gamma ^A \partial _A \psi _r + \frac{1}{2} \cot \theta \gamma ^2 \psi _r +\gamma ^u \partial _u M  \psi _r  + \gamma ^{u } \partial _{u } \psi _r ^{-2}  - \gamma ^A \psi _A  =0 
\end{flalign}

\end{itemize}

We see that the leading order components of the field satisfy $  \phi _{\mu} =  \phi _{\mu} (x^A) + \gamma ^{u(0)}\varphi _{\mu} (u, x^A)$. We also notice that the equations for $\psi _A$ are quite complicated to solve. However, we don't need to solve for $\psi _{\theta}$ and $\psi _{\phi }$ individually. From the expression for the charge \eqref{Q} it is clear that what we need is $\gamma ^A \psi _A$. The gauge condition $\gamma ^{\mu } \psi _{\mu}$ can help us here. We demand that $u$ dependent and independent part vanish separately:

\begin{flalign} \label{gauge2}
\gamma ^{u(0)} \varphi _{u}   + \gamma ^{r (0)} \varphi _{r}   + \gamma ^{A(1)} \varphi _{A }= 0\\
\gamma ^{r (0)} \gamma ^{u(0)} \phi _{r}  + \gamma ^{A {1}} \gamma ^{u(0)} \phi _{A}  = 0
\end{flalign}

We see that instead of looking at $\gamma ^A \psi _A$, we can look at the equations for $\varphi _r$, $\varphi_u$ and $\phi _u$, which are much simpler. We will do this in the following way - we will extract the $u$ dependent and $u$ independent part of equations \eqref{r2} and \eqref{u2} set them separately to be equal to zero.

We begin by studying $\varphi _u (x^A)$. It's equation of motion is,

\begin{flalign}
 \gamma ^{A(-1)} \partial _A \varphi _u  + \frac{1}{2} \cot \theta \gamma ^2 \varphi _u  = 0
\end{flalign}

This is the exact same equation as for $\rho(x^A)$ and therefore it has the same solutions.

We move on to investigate $\varphi _r$.  Multiplying \eqref{r2} by $\gamma ^{u(0)}$ and replacing $ \gamma ^{u(0)}  \gamma ^{A {1}}\varphi _{A}  $, by $-  \gamma ^{u(0)}  \gamma ^{r (0)} \varphi _{r} $ we get:

\begin{flalign}
\gamma ^{u (0)} \left(\gamma ^A \partial _A \varphi  _r + \gamma ^1 \varphi _r +  \frac{1}{2} \cot \theta \gamma ^2 \varphi  _r \right) =0
\end{flalign}

The above equation means one of two things. Either $ \varphi _r $  is a covariantly constant spinor or the thing in brackets is in the image of $\gamma ^{u(0)}$. When this happens every component of $\varphi _r$  is a linear sum of spin $\pm 1/2$ spherical harmonics, with some conditions on the coefficients. The details of this calculations are quite long and are therefore, given in the appendix.

We finally look at $ \gamma ^{u(0)}\phi _r(u, x^A)$. This equation for it is,

\begin{flalign}
&\gamma ^{u(0)} \left( \gamma ^{A(-1)} \partial _A\phi _r(u, x^A)+ \frac{1}{2} \cot \theta \gamma ^2  \phi _r(u, x^A) + \partial _u M  \varphi _r +\partial _{u } \psi _r ^{-2} -  \gamma ^{1} \phi _{r} \right)  = 2 \gamma ^0 \gamma ^1 \phi _{r}\\
& \gamma ^{u(0)} \gamma ^0 \gamma ^1 \phi _r = 0 \,  \Leftrightarrow \, \gamma ^{u(0)}  \phi _r   = 0
\end{flalign}

Thus we see that $u$ dependent part of $\psi _r ^{-1}$ and consequently of $\psi _A ^0$ vanish.

\subsection{Result}

To sum up we have,

\begin{flalign}
&\gamma ^{A(-1)} \psi _A ^{0} =  - \gamma ^{u(0)} \varphi _{u}   - \gamma ^{r (0)} \varphi _r \quad \epsilon ^{-1}  = \rho  + \gamma ^{u(0)} \beta \\
&  \gamma ^{u(0)} \varphi _{u} = \frac{\left( \cot \theta /2 \right) ^{m}}{\sqrt{\sin \theta }}  \begin{pmatrix} i A \sin m \phi  \\ i A \sin m\phi  \\-B \cos m \phi  \\ B  \cos m \phi \end{pmatrix} \quad \varphi _r = \begin{pmatrix} \zeta \\ \chi \end{pmatrix} \\
& \gamma ^{u(0)} \beta= \sum _{lm } \begin{pmatrix}   \Im \{ a(u) _{1/2} ^{lm} \, _{1/2} Y_{lm } \} \\   \Im \{ a(u) _{1/2} ^{lm} \, _{1/2} Y_{lm } \} \\ -\Re \{ b(u) _{-1/2} ^{lm} \, _{-1/2} Y_{lm } \} \\  \Re \{ b(u) _{-1/2} ^{lm} \, _{-1/2} Y_{lm } \} \end{pmatrix}  \quad \rho =  \frac{\left( \cot \theta /2 \right) ^{m}}{2 \sqrt{\sin \theta }}  \begin{pmatrix}i C \sin m\phi \\  i D \sin m \phi  \\ F \cos m \phi \\
G \cos m \phi  \end{pmatrix}
\end{flalign}
where the expression for $\zeta$ and $\chi$ are quite long and are in the appendix.

The charge is,

\begin{flalign}
\mathbf{Q} [\epsilon ] = & \int _{S^2 } d\theta d\phi \, \sin
\theta \left[\bar{\varphi } _r \gamma ^0 \epsilon ^0 +
\bar{\varphi } _u \gamma ^{u(0)}\gamma _5 \gamma _2 \gamma _3
\rho - \frac{\theta _{\ast} }{2} \bar{\varphi } _r \gamma ^0
\gamma _5 \epsilon ^0 + \frac{\theta _{\ast} }{2} \bar{\varphi }
_u \gamma ^{u(0)} \gamma _2 \gamma _3 \rho \right]
\end{flalign}

\section{Conclusion}

We just discovered and infinite number of electric and magnetic charges at null-infinity for thefree massless Rarita-Schwinger field. We were able to compute those charges explicitly. They exhibit some peculiar properties. Firstly, we notice that even though we have non-vanishing Noether magnetic charges, the symplectic structure of the theory is unaltered by the presence of the new boundary term. Furthermore, the charges can have an arbitrary time dependence. As mentioned in the introduction, the next step is to extend this work in supergravity. Hopefully, the interplay of the spinor magnetic charges and the gravitational dual charges will shed new light on the present work.
\appendix

\section{Conventions}\label{conv}

The indices latin $a$ are tangent space indices, and the greek indices $\mu$, refer to space time.
The matrices  $\gamma _a$, $a=0..3$ are the Dirac matrices and $\gamma _{\mu } =  \gamma _a e^a _{\mu }$. The covariant derivative is written with $\nabla$. Its action on spinor-vectors is $\nabla_{\mu } \psi _{\nu } = \partial _{\mu } \psi _{\nu } + \frac{1}{4} \omega  _{\mu ab} \gamma ^{ab} \psi _{\nu } - \Gamma ^{\lambda } _{\mu \nu  } \psi _{\lambda } $. On the frame field it is $\nabla _{\mu } e_{\nu } ^a = 0$, because the space-time is without torsion. Furthermore  $\gamma ^{ab} = \frac{1}{2} [\gamma ^a, \gamma ^b]$, $\gamma ^{abc} =\frac{1}{2} \{ \gamma ^{ab}, \gamma ^c \}$ etc. The particular form of gamma matrices and the charge conjugation  $C$ matrix we have picked is:

\begin{flalign}
\gamma ^0  =-i \begin{pmatrix} 0 &\mathbb{1} \\ \mathbb{1} & 0 \end{pmatrix} \quad \gamma ^j  = -i \begin{pmatrix} 0 &  \sigma ^j   \\ - \sigma ^j & 0 \end{pmatrix} \quad  \gamma ^5  =i \gamma _0 \gamma _1 \gamma _2 \gamma _3= \begin{pmatrix}- \mathbb{1} &  0  \\ 0 & \mathbb{1} \end{pmatrix}  \quad C = -i\begin{pmatrix} 0 &\mathbb{1} \\  - \mathbb{1} & 0 \end{pmatrix} 
\end{flalign}

\section{Equation of motion for $\varphi _r$}

The equation for $\varphi _r$ is:

\begin{flalign}
\gamma ^{u(0)}\left( \gamma ^A \partial _A \varphi  _r + \gamma ^1 \varphi _r +  \frac{1}{2} \cot \theta \gamma ^2 \varphi  _r \right) = 0
\end{flalign}

We can write the spinor as $\varphi _r= \begin{pmatrix} \zeta \\ \chi \end{pmatrix}$. We set thing in brackets to be  equal to a generic spinor in the image of $\gamma ^{u(0)}$:

\begin{gather}
 -i\partial _{\theta }\chi _2 + \frac{1}{\sin \theta } \partial _{\phi } \chi _1 - \frac{i}{2} \cot \theta \chi _2 =- \chi _2  + \alpha \\
 i \partial _{\theta }\chi _1 - \frac{1}{\sin \theta } \partial _{\phi } \chi _2 + \frac{i}{2} \cot \theta \chi _1 =   -\chi _1 +\alpha
\end{gather}

First we look at the case, where $\alpha=0$. We pose $\chi _1 =e^{- i \frac{\theta }{2 }} f(\phi )$ and $\chi _2 =e^{ i \frac{\theta }{2 }} h(\phi )$:

\begin{gather}
 e^{- i \frac{\theta }{2 }} \partial _{\phi } f - \frac{i}{2} e^{ - i \frac{\theta }{2 } } h = 0 \quad 
e^{ i \frac{\theta }{2 }} \partial _{\phi } h  -\frac{i}{2} e^{ i \frac{\theta }{2 } } f = 0\\
\Rightarrow \partial _{\phi } h  = \frac{i}{2} f \quad \partial _{\phi } f  = \frac{i}{2} h
\end{gather}

Setting $h=a e^{i \frac{\phi }{2}} + b e^{ - i \frac{\phi }{2}} $ and $f=c e^{i \frac{\phi }{2}} + d e^{ - i \frac{\phi }{2}} $ we have $a=c$ and $b=-d$. Now let's consider what happens if $\alpha \neq 0$:

\begin{flalign}
- \bar{\eth} _{1/2}(\chi _2 - \chi _1) = i (\chi _2 + \chi _1) - 2i \alpha \\
-\eth _{-1/2} (\chi _2 + \chi _1) = i (\chi _2 - \chi _1)
\end{flalign}

This has solution if we set $\chi _2 - \chi _1 = \sum a^{lm} _{1/2}  Y _{lm}$, $\chi _2 + \chi _1 = \sum b^{lm} _{-1/2}  Y _{lm}$ and $\alpha   =  \sum \alpha ^{lm} _{-1/2} Y _{lm}$. The equations, satisfied by these coefficients are:

\begin{flalign}
&a ^{lm} (l+1/2)=i b ^{lm} -2 i \alpha ^{lm} \quad i b^{lm}(l+1/2 )=  a ^{lm}\\
\Rightarrow \, & b^{lm} (l- 1/2) ^2 = -2 \alpha ^{lm}
\end{flalign}

We see that the solution we found for when $\alpha = 0$ can be intuitively interpreted as the $1/2$ mode spherical harmonic.

We now solve quickly for $\zeta$ which has almost the exact same equations.

\begin{flalign}
&- \bar{\eth} _{1/2}(\zeta _2 - \chi _1) = i (\zeta _2 + \zeta _1)  \\
&-\eth _{-1/2} (\zeta _2 + \zeta _1) = i (\zeta _2 - \zeta _1) - 2i \upsilon
\end{flalign}

We set again $\zeta _2 - \zeta _1 = \sum c^{lm} _{1/2}  Y _{lm}$, $\zeta _2 + \zeta _1 = \sum d^{lm} _{-1/2}  Y _{lm}$ and $\upsilon   =  \sum \upsilon ^{lm} _{-1/2} Y _{lm}$, which gives

\begin{flalign}
& c ^{lm} (l+1/2)=i d ^{lm}  \quad i d^{lm}(l+1/2 )=  c ^{lm} -2  \upsilon ^{lm}
\end{flalign}

The case where $\upsilon = 0$ is exactly the same as before. Otherwise the difference is in the coefficients.
The final thins is to remember to impose the Majorana condition $\zeta ^{\ast } = \zeta $ and $\chi  ^{\ast } = - \chi$. In the end chiral spinors are:

\begin{flalign}
& \zeta  = i \begin{pmatrix}    c \sin \frac{\phi -  \theta }{2} - d  \sin \frac{\phi +  \theta }{2}  \\   c \sin \frac{\phi -  \theta }{2} + d \sin \frac{\phi +  \theta }{2}  \end{pmatrix} +  \frac{1}{2} \sum _{lm} \Im \begin{pmatrix}  -  c^{lm} \, _{1/2} Y _{lm} + d^{lm} \, _{-1/2}  Y _{lm}\\   c^{lm}\, _{1/2} Y_{lm} + d^{lm}\, _{-1/2}  Y _{lm} \end{pmatrix} \\
& \chi  = \begin{pmatrix}   a \cos \frac{\phi -  \theta }{2} + b  \cos \frac{\phi +  \theta }{2}  \\  a \cos \frac{\phi -  \theta }{2} + b  \cos \frac{\phi +  \theta }{2}  \end{pmatrix} +  \frac{1}{2} \sum _{lm} \Re \begin{pmatrix}  -  a^{lm} \, _{1/2} Y _{lm} + b^{lm} \, _{-1/2}  Y _{lm}\\   a^{lm}\, _{1/2} Y_{lm} + b^{lm}\, _{-1/2}  Y _{lm} \end{pmatrix}
\end{flalign}

\section{Spin connection explicitly up to second order}

\begin{flalign}
\mathbf{\omega _{01} } = &\left( \frac{2\partial _u M }{r}  +  \frac{M-6M\partial_u M}{r^2} + \partial _u \beta \right)  \mathbf{e^0} -\left( \frac{2\partial _u M }{r}  -  \frac{6M\partial_u M}{r^2} + \partial _u \beta \right)\mathbf{e^1} \\
& - \left( E_{iA}U^A  + 2 \frac{E^{A(0)} _i \partial_A \beta }{r} \right) \mathbf{e^i} 
\end{flalign}

\begin{flalign}
\mathbf{\omega _{0i}}  = \left( -\frac{E^{A(0)} _i \partial _A M}{r^2} \right)\mathbf{e^0}  + \left( \frac{ 2 E^{A(0)} _i \partial _A M}{r^2} - E_{iA}U^A \right)\mathbf{e^1} + \left(  -2\left( \delta _{ij } \partial _u C_{Ai}\right)  + ...\right)\mathbf{e^j }
\end{flalign}

\begin{flalign}
\mathbf{\omega _{1i}} = &\left( E_{iA} U^A  + rE_{iA}\partial _r U^A  + 2\frac{E^A _i \partial _A M}{r^2}\right) \mathbf{e^0} + \left(  2\frac{E^A _i \partial _A M}{r^2} \right) \mathbf{e^1} + \left( \delta _{ij}\frac{1}{r} (1-\beta)  - \delta _{ij}\frac{M}{r^2}  \right) \mathbf{e^j}
\end{flalign}

\begin{flalign}
\mathbf{\omega_{ij}} = & 2 \left( 2  \partial _{[A}  E ^{(0)} _{B]i}U^AE^{B(0)} _j  \right) \mathbf{e^0} -2 \left( 2  \partial _{[A}  E ^{(0)} _{B]i}  U^A E^{B(0)} _j  \right) \mathbf{e^1}\\
& + \left( \frac{4}{r} \partial _{[A}  E _{B][i}  E^A _{j]} E^B _k -  \frac{2}{r} \partial _{[A}  E _{B]k}  E^A _i E^B _j \right)\mathbf{e^k}
\end{flalign}

\section{Finiteness of energy and momentum for the field} \label{T}

The energy and momentum for a field are defined as the integral of different components of the stress-energy tensor over a space like surface. In order for this to be finite in four dimension we need \protect $T^{\alpha  u } \sim  O(r^{-2})$. On shell the expression for the stress-energy tensor of the free massless Rarita-Schwinger field is,

\begin{flalign}
T_{\mu \nu  }& = - \frac{2}{\sqrt{g}} \frac{\delta S} {\delta g^{\mu \nu }} \\
 \frac{1}{2} T^{\alpha \beta} & \approx \frac{1}{\sqrt{g}}e ^{\mu \nu  \sigma ( \alpha } \left( \bar{\psi } _{\mu }  \gamma _5  \gamma _{\sigma } \left[  \nabla _{\nu } \psi ^{\beta )} -   \nabla ^{\beta)} \psi _{ \nu } \right] + \theta  _{\ast} \bar{\psi } _{\mu }  \gamma _{\sigma } \left[  \nabla _{\nu } \psi ^{\beta )} -   \nabla ^{\beta)} \psi _{ \nu } \right] \right) \\
 T^{u \alpha } & = \frac{1}{\sqrt{g}}  e ^{\mu \nu  \sigma u } \left( \bar{\psi } _{\mu }  \gamma _5  \gamma _{\sigma } \left[  \nabla _{\nu } \psi ^{\alpha } -   \nabla ^{\alpha } \psi _{ \nu } \right] +\theta  _{\ast} \bar{\psi } _{\mu }  \gamma _{\sigma } \left[  \nabla _{\nu } \psi ^{\alpha } -   \nabla ^{\alpha } \psi _{ \nu } \right] \right) \\
& \Rightarrow \bar{\psi } _{[r }  \gamma _5  \gamma _{\theta }  \nabla _{\phi ] } \psi ^{\alpha}  \sim O(1)
\end{flalign}

This is satisfied, by the conditions we have set -\protect $\psi _r \sim \psi _u \sim O(r^{-1})$ and \protect $\psi _A  \sim O(1)$.

Note that, by the equations of motion of Rarita-Schwinger field, \protect $T^{\alpha } _{\, \, \alpha }  =0$. Furthermore, because of the same arguments we used to prove the invariance of the action under gauge transformation, the stress-energy tensor is has vanishing divergence.

\begin{flalign}
 \nabla _{\alpha } T^{\alpha}_{ \ \ \beta } &\approx  \bar{\psi } _{\mu }  \gamma _5  \gamma ^{\mu \nu  ( \alpha} \nabla _{\alpha }\left[  \nabla _{\nu } \psi _{\beta )} -   \nabla_{\beta)} \psi _{ \nu } \right] +\theta  _{\ast}\bar{\psi } _{\mu }  \gamma ^{\mu \nu  ( \alpha} \nabla _{\alpha }  \nabla _{\alpha }\left[  \nabla _{\nu } \psi _{\beta )} -   \nabla _{\beta)} \psi _{ \nu } \right]  \\
& =\bar{\psi } _{\mu }  \gamma _5  \gamma ^{\mu \nu  ( \alpha} \nabla _{\alpha } R_{\alpha [\nu | ab} \gamma ^{ab} \psi _{ | \beta ] }+\theta  _{\ast} \bar{\psi } _{\mu}  \gamma ^{\mu \nu  ( \alpha} \nabla _{\alpha }  R_{\alpha [\nu | ab} \gamma ^{ab} \psi _{ | \beta ] } = 0 \\
\end{flalign}

\section*{Acknowledgments}

The author thanks the contribution of her supervisor Malcolm Perry, in motivating, discussing and reviewing
the contents of this paper, Jonathan Crabbé, Rifath Khan, Filipe Miguel and Gon\c{c}alo Regado for fruitful physics discussions. The author is jointly funded by the University of Cambridge, the Cambridge Trust and King’s College.

\end{document}